\documentclass[aip,cha,reprint,graphicx,amsmath,amssymb]{revtex4-1}
\usepackage[pdftex]{graphicx} 
\usepackage{CJKutf8}
\usepackage[T1]{fontenc}
\usepackage{mathptmx}
\usepackage{hyperref}
\newcommand{\argmax}{\mathop{\rm arg~max}\limits}

\usepackage{enumerate}

\draft 
 
\begin{document}


\title{Functional differentiations in evolutionary reservoir computing networks}

\author{Yutaka Yamaguti}
\email{y-yamaguchi@fit.ac.jp \\This article has been submitted to Chaos. }
\affiliation{Faculty of Information Engineering, Fukuoka Institute of Technology, Fukuoka, 811-0295, Japan}

\author{Ichiro Tsuda}%
\affiliation{ 
  Chubu University Academy of Emerging Sciences, Chubu University, Kasugai, Aichi, 487-8501, Japan
}%

\date{\today}

\begin{abstract}
  We propose an extended reservoir computer that shows the functional differentiation of neurons. The reservoir computer is developed
  to enable changing of the internal reservoir using evolutionary dynamics, and we call it an evolutionary reservoir computer. 
  To develop neuronal units to show specificity, depending on the input information, the internal dynamics should be controlled
  to produce contracting dynamics after expanding dynamics. Expanding dynamics magnifies the difference of input information, while
  contracting dynamics contributes to forming clusters of input information, thereby producing multiple attractors.
  The simultaneous appearance of both dynamics indicates the
  existence of chaos. In contrast, sequential appearance of these dynamics during finite time intervals
  may induce functional differentiations. In this paper, we show how specific neuronal units are yielded in the evolutionary reservoir computer.
\end{abstract}


\maketitle 

\begin{quotation}
  One of the universal characteristics of the brain is functional differentiation, where each elementary unit of the brain, that is,
  a neuron or a neuron assembly, plays a role in a respective specific function. The emergence of functional differentiation
  in the brain depends not only on gene expressions but on self-organization with constraints. Here, constraints may stem from
  physical, chemical, and even informational factors that develop fetal brain in its developmental process.
  In this paper, we propose an artificial neural network that realizes functional differentiation. In particular,
  we focus on computation ability of an extended reservoir computer, introducing evolutionary dynamics, by which
  parameters such as the connection weights of an internal network can be changed and optimized for a given constraint.
  In such an evolutionary reservoir computer, we found that neurons differentiated to respond to specific
  input stimulations, where we used spatial and temporal patterns as representations of visual and auditory stimulations, respectively.
  In this dynamic development, the network topology changed from a random network to a feedforward network including
  feedback connections. The development of this type of network structure is consistent with animal and human cortical local
  network structures. Although the structure of the present system and its dynamical rule are different from those in R\"ossler's optimization system,
  both converged dynamics are similar, in the sense that an optimized solution adapts to the environment.
\end{quotation}

\section{Introduction}
\label{introduction}
Functional differentiations have occurred
in animal and human brains in biological evolution associated with structural changes of the
brain networks \cite{sporns2016,Glasser2016,treves1996,sharma2000}. Furthermore, in the cerebral cortex, more than 180 different areas
were recently identified using modern imaging technology, where
assemblies of multiple modules temporarily emerged depending on given tasks via cooperative dynamics across such assemblies,
and this process is called functional parcellation \cite{Glasser2016}.
In contrast, similar correlated behaviors between modules have also been
observed in resting states,
in association with transition dynamics of neural activity between functional areas \cite{biswal1995,greicius2003,fox2005}.
This is known as transitions between default modes.
The questions have been presented regarding why and how functional differentiation
and/or functional parcellation occurred. However,
only a few theoretical studies have been proposed thus far\cite{malsburg1973,amari1980,kohonen1982},
 and the underlying basic principle is still unclear.
In this paper, we use the term functional differentiations as a generic term expressing both functional differentiation and functional parcellation.

Clarifying the computational principle of functional differentiations leads not only to a fundamental
understanding of the brain's functions but also to a promising direction of research for artificial
neural networks that possess a flexible and adaptive structure. The study of such artificial networks may provide a better design
for artificial intelligence and robotics\cite{le2013,asada2001}. It is also known that genetic factors do not entirely
determine the form of differentiations\cite{sharma2000}. In the present paper, we treat functional differentiations from the perspective of
self-organization in the brain, whereby the computational principle of functional differentiations can be clarified.

We have proposed the self-organization with constraints concept for describing functional
differentiations in the brain and studied several mathematical
models \cite{tsuda2016}. In this study, we observed the emergence of 
system components (i.e., elements) caused by a given constraint that acts on a whole system.
In Ref. \onlinecite{yamaguti2015}, using a genetic algorithm, we studied how heterogeneous modules develop in the networks consisting of coupled phase oscillators of the modified Kuramoto model\cite{kuramoto1984}.
In that study, the constraint was provided by maximizing the bidirectional information transmission between
two subnetworks. Starting from the random networks of phase oscillators, two functionally
differentiated modules evolved. The emergence of the differentiated modules from random and homogeneous networks may be caused by  
symmetry breaking via chaotic behaviors of the network.

As a further extended study about self-organization with constraints, here we treat the functional differentiation of networks in terms of a biologically inspired model that can operate cognitive tasks.
We also investigate what constraints can be appropriate for the emergence of functional differentiations.
The present model is based on reservoir computers\cite{maass2002,jaeger2004,dominey1995}.

Reservoir computers were proposed as a kind of recurrent neural network (RNN) and were considered as computational
models of cerebellum cortex\cite{yamazaki2007} and neocortex\cite{enel2016} because of similar network structures
to microcircuits in the brain, such as random and sparse connections.
Unlike other RNNs and deep neural networks, only synaptic weights to output units (called readout units) are modified, and recurrent synaptic weights remained
unchanged throughout the learning process. This scheme drastically speeds up the supervised learning process compared with back-propagation-based methods.
Furthermore, fully developed chaos in the internal network produces poor performance, but several studies have reported that the "edge of chaos", or weakly chaotic states of the internal network, can bring about effective performance for given tasks \cite{bertschinger2004, sussillo2009, hoerzer2014}.
 However, other studies argue that the importance of the edge of chaos does not always hold \cite{laje2013, carroll2020}. 
 It may be more essential for certain tasks, such as short-term memory, that the system state is at the "edge of stability" \cite{carroll2020, yamane2016}, where the largest Lyapunov exponent of an attractor is near zero.
 Examples of information processing in neural networks using such neutral stability have been discussed in, for example, Refs. \onlinecite{mante2013, yamane2016}.

 As stated in the following paragraphs, we treat optimization problems by introducing evolutionary dynamics. Using an optimization method, the complexity of network dynamics is reduced by the appearance of fixed-point states in the case without inputs, and then the network dynamics is easily entrained by the input dynamics, thereby adapting to environments.
Coupled optimizers of R\"ossler\cite{rossler1987} provide a typical example of the appearance of fixed-point dynamics in relative dynamic behaviors between
controlled and controlling systems, while overall dynamics appears to be weakly chaotic states.

In conventional studies of reservoir computers (RCs), the internal network, which plays a role in a reservoir, is constructed by 
random connections of neuronal units. However, recently, the study of different types of network topology such as small-world 
network is attracting attention because such a network provides effective performance\cite{kawai2019,park2019}. 
Accordingly, we here study the functional differentiations of RCs that can perform multiple tasks, by exploring an optimized network. 
In this respect, we propose an extended model of RCs, by adopting genetic algorithms to such an internal recurrent network, which we call
evolutionary RCs (ERCs). In an optimization problem in an ERC, we combined an evolutionary algorithm, which modifies the structure of recurrent networks, 
with a supervised learning algorithm, which changes the output weights.

In the present study, we consider a network that receives combined signals coming from multiple sensory modalities, such as visual and auditory sensations, which were encoded by spatial and temporal patterns, respectively. The task of the network is to separate the spatial and temporal patterns from the combined input signals. The performance of the network is evaluated after determining the output weights. Using an evolutionary algorithm,  we investigate a network structure that realizes high task performance.  From the perspective of functional differentiations, the internal structure of the developed network was studied, using mutual information analysis.

The organization of the paper is as follows. In Section \ref{sec-model}, we describe the mathematical model, tasks, genetic algorithm, and the method of network analysis.
In Section \ref{sec-results}, we show how ERCs can solve the given tasks by changing internal network structures. The information mechanism of functional differentiations in neuronal units of the networks is clarified using mutual information analysis, and the parameter dependence on the functional differentiations is also shown. Section \ref{sec-discussion} is devoted to summary and discussion.

\section{Model and Method}
\label{sec-model}
\subsection{Network architecture}

\begin{figure}
  \includegraphics[width=8.5cm]{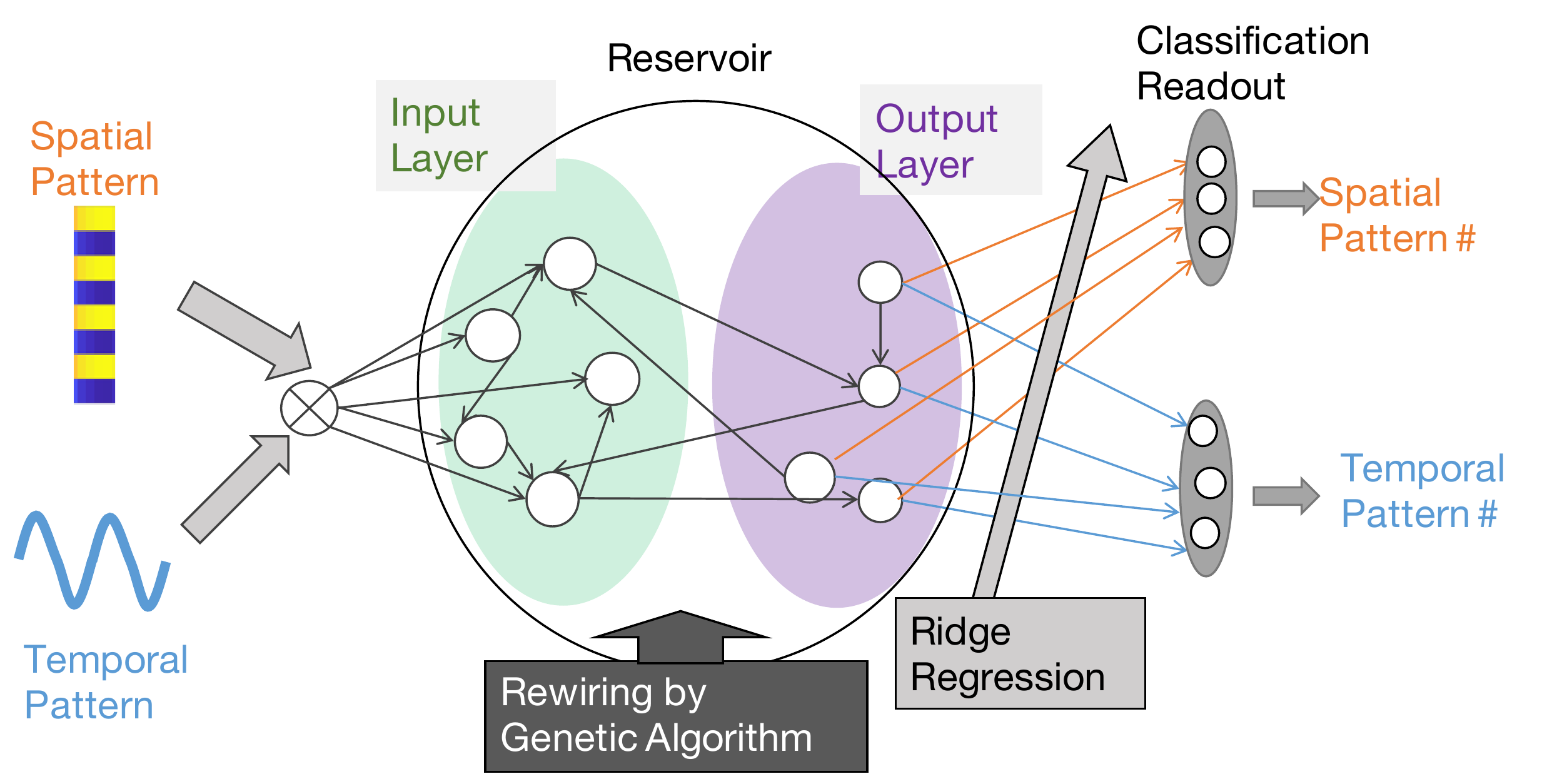}
  \caption{Network architecture. Two kinds of input patterns are superimposed as an input. The internal network is separated into two layers, input and output layers. 
    Output signals are represented in spatial and temporal readout layers. \label{fig-network}}
\end{figure}

We extended RCs \cite{dominey1995,maass2002,jaeger2004}, adopting the optimization principle via evolutionary dynamics
to the internal random recurrent networks, called ERCs.
We employed a continuous-valued, discrete-timestep model of ERC.
The network consists of $N$ neurons, whose dynamics is described by
\begin{eqnarray}
  x_{i}(t+1) =&& (1-\alpha_{i}) x_{i}(t) \nonumber\\
   && +\alpha_{i} \tanh \left( \sum_{j}{w_{ij}x_j(t) + w_{i0} + \sum_{k} w^{(\text{in})}_{ik} I_{k}(t)} \right) \nonumber\\
   && +\xi_{i}(t), \label{eq-esn}
\end{eqnarray}

where $x_{i}(t)$ is a state of the $i$-th neuron at time $t$, 
 $w_{ij}$  and $w^{(\text{in})}_{k}$ are weights of recurrent connections and weights of connections to an input signal $I_k(t)$, respectively, and $w_{i0}$ is a constant term. In addition, $\alpha_i$ represents a decay constant, and $\xi_{i}(t)$ is a noise term. 
The initial state of $x_{i}$ was randomly taken from a uniform distribution over $[-0.5, 0.5]$.
 The network consisting of these neurons was constructed with sparse connections, such that only 10\% of synaptic connections on the network were nonzero.
 
 Because of effective computations on differentiations, we used the following additional structure of internal networks. 
 The internal networks were divided into two groups of neurons that
 construct an input and an output layer within internal networks, unlike a standard RC that is constructed by homogeneous random networks.
 Neurons in the input layer receive input signals but have no direct connections to output units, called readout units. In contrast, neurons in the output layer do not receive input signals directly but have direct connections to the readout units (Fig. \ref{fig-network}).

The model was trained to perform the discrimination task of two different kinds of patterns: simultaneous discrimination of 
spatial and temporal patterns. Correspondingly,
the model possesses two distinct output vectors, spatial $y^{\text{(sp)}}$ and temporal $y^{\text{(temp)}}$ vectors.
Each output vector was determined by a linearly weighted sum of the internal states, as shown in the following equation.
 \begin{equation}
 y^{\text{(*)}}_{i}(t) = \sum_{j} w^{\text{(*)}}_{ij} x_{j}(t),
\end{equation} 
where \text{(*)} indicates spatial or temporal patterns, $i=1, ...,L$ are indices of readout units and $w^{\text{(*)}}_{ij}$ are weights for readout.
In the present simulation, output weights are determined, using ridge regression. 
For input patterns denoted by an asterisk,
the following loss function was minimized:
\begin{equation}
   Q^{\text{(*)}} = \frac{1}{T_{\text{tr}}L} \sum_{t,i}{ (y_{i}^{(*)}(t)- p_{i}^{(*)}(t))^2 }+\mu \sum_{i,j} { (w^{\text{(*)}}_{ij})^2}, \label{eq-loss}
\end{equation}
where the first term indicates the loss stemming from the differences between output and target patterns $p_{i}^{(*)}(t)$ 
during a training period of length $T_{\text{tr}}$ and
the second term indicates the loss from the weight $L^{2}$ regularization. 

\subsection{Task setting}
\subsubsection{Separation Task}
To represent a simultaneous input of visual and auditory patterns, we considered the product of spatial and temporal patterns as inputs.
     The input was represented by an $N_{\text{in}}$-dimensional signal $I(t)$. We predefined finite sets of
spatial patterns $\left\{ a^{(\bar{l})} \right\} \ (\bar{l}=1,... , L) $ and temporal patterns $\left\{ b^{(\bar{m})}(t) \right\} (\bar{m}=1,... , M)$.
Then, for every time period $\tau$, we multiplied randomly selected temporal and spatial patterns to generate the input signal.
Spatial patterns $a^{(\bar{l})} \in \mathbb{R}^{N_{\text{in}}}$ represent static components in sensory inputs, which we realized as square waves by the following equation:            

\[
 a^{(\bar{l})}_{k}= \begin{cases} 
    -1, & \text{if \ }  2^{\bar{l}-1} (k-1)/ N_{\text{in}} \text{\ \  (mod 1)}  < 1/2 \\
    1, & \text{otherwise}.
    \end{cases}.
\] 
Temporal patterns $b^{(\bar{m})}(t) \in \mathbb{R}$ were represented as one-dimensional sinusoidal signals:
\[ b^{(\bar{m})}(t)=\cos(2 \pi f^{(\bar{m})} t),\]
where $f^{(\bar{m})}=1/2^{\bar{m}+2}$.
Here, let  $l(t)$ and $m(t)$ be the indices of spatial and temporal patterns at time $t$, respectively.
Thus, when the $l(t)$-th spatial pattern and $m(t)$-th temporal pattern at time $t$ are presented, 
the input signal of the $k$-th unit, $I_k(t)$ in Eq. ~(\ref{eq-esn}) is described by $I_k(t)  = a^{(l(t))}_{k} b^{(m(t))}(t)$.
The combination of spatial and temporal patterns, $(l(t), m(t))$, was randomly switched for each $\tau=64$ step.
In the present paper, the computational results are shown in the case of $L=M=3$.

The task of ERCs is to discriminate spatial and temporal patterns 
and classify each pattern correctly at the readout units.
The network was trained such that the $l$-th unit of the spatial and the $m$-th unit of the temporal readout units exhibit ones, 
and the other neuronal units exhibit zeros.
Considering the delay from the input to the output layers, the input pattern of four steps ago was set as a teacher signal, i.e., 
$p^{(\text{sp})}_{i}(t)$ in Eq.~(\ref{eq-loss}) takes $1$ if $i = l(t-4)$ and $0$ otherwise. The same rule was applied to $p^{(\text{temp})}_{i}(t)$.
To perform this kind of task, the network needs
short-term memory and nonlinear operations. 
The accuracy of network learning was calculated as a proportion of time such that the readout unit taking the maximum value
$\argmax_{i} y^{(*)}_{i}(t) $ at each time coincides with the correct index $\argmax_{i} p^{(*)}_{i}(t) $. 

\subsubsection{Combination Task}
For comparison with the presented performance of the network, we tried to evolve the network with another task called the combination task. The task was
detecting the specific combinations of spatial and temporal inputs. 
In the previous task, processing two information sources after separating such sources was necessary,
while the present task needs processing of combined information of two sources. We used the same input patterns as the previous task.
The same number of readout units as the spatial (and temporal) readout units in the separation task ($L=M$) was used for fair comparison between these two conditions.
In supervised learning at readout units $y^{(\text{combi})}_{i}(t)$, $(i=1, ..., L)$, we changed the connection weights to detect some specific combinations of spatial and temporal patterns. 
In more detail, let $S_{i}$ $(i=1,.., L)$ be a set of combinations of spatial and temporal patterns, which the $i$-th readout unit should detect. Furthermore, each combination $(\bar{l},\bar{m})$ $(\bar{l}=1,\cdots, L, \bar{m}=1,\cdots, M)$
is included in either set of $S_{i}$. An example of this allocation of a combination of patterns to readout units is shown in Fig. \ref{fig-ex-combi}. Considering the time delay in a similar manner of the case of a separation task,
the teacher signals $p^{(\text{combi})}_{i}(t)$ $(i=1,\cdots, L)$ takes 1 only when $(l(t-4),m(t-4)) \in S_{i}$, and otherwise 0.
In the evolution experiment, the network size and other system parameters were the same as in the separation task.
The combination of $S_{i}$ was randomly allocated in each generation.

\subsection{Genetic algorithm}
\label{subsec-evolution}
For the presented tasks, reservoir computing networks with random internal networks, which belong to conventional RCs, did not show good performance, even after
the output weights were learned using the ridge regression. This implies that the structure of internal networks must be changed for such tasks.
This is a reason why we introduced a genetic algorithm for changing the internal networks and extended conventional RCs to evolutionary ones.
For the first time, we prepared an initial ensemble of randomly connected networks, 
and second, we performed the following procedures (1) to (3) for each network.
\begin{enumerate}[(1)]
  \item Collection of data for learning. Input a pattern sequence consisting of spatial and temporal patterns, and continue to update internal states according to Eq. (\ref{eq-esn}), and discard the first transient data.
  \item  Learning period. Perform supervised learning of the readout weights with ridge regression, using the internal states and teacher signals collected in (1).

  \item  Test period. Update the states of both internal and readout units, and then calculate the root mean square of errors between teacher signals and outputs for both spatial and temporal patterns. Then, evaluate Eq.(\ref{eq-objective}).
\end{enumerate}
For all networks, the following procedure (4) was performed. 
\begin{enumerate}[(4)]
\item Evolutionary period. Among all networks, select networks with smaller errors for which the next generation of the networks were produced using mutation and crossover.
\end{enumerate}

In the present simulation, the targets for mutation and crossover are the weights of recurrent connections and decay constants, $\alpha_{i}$.
Mutation of the weights of recurrent connections was performed, randomly reconnecting $4$\% of the total connections and  
adding Gaussian noise, $N(0, \sigma_w^2)$ in a randomly selected 40\% of the weights of all internal connections.
Decay constants 
$\alpha_{i}$ were also changed by randomly adding Gaussian noise, $N(0, \sigma_{\alpha}^2)$, where the upper and lower cutoffs of $\alpha_{i}$ were commonly given as
[$\alpha_{\text{low}}, \alpha_{\text{high}}]$. 
As for crossover, we randomly selected two survived networks and constructed one new network, where a half connection of a new network stems from a randomly chosen half connection of one survived network and the other half connection from the other survived network. 
For each generation, we repeated (1) to (4), and the selected networks show high adaptability to the given task.
As a constraint for evolutionary dynamics, the following loss function was adopted, which indicates the summation of mean squared errors of the activity of all readout units, based on the teacher signals for both spatial and temporal patterns. Here, an average was taken over both of 
all readout units and a test period of length $T_{\text{te}}$.
\begin{eqnarray}
 Q^{\text{(\text{evo})}} = &&\frac{1}{T_{\text{te}}M} \sum_{t,i}{ (y_{i}^{(\text{sp})}(t)- p_{i}^{(\text{sp})}(t))^2 } \nonumber\\
                           &&+\frac{1}{T_{\text{te}}L} \sum_{t,i}{ (y_{i}^{(\text{temp})}(t)- p_{i}^{(\text{temp})}(t))^2 }. \label{eq-objective}
\end{eqnarray}

\subsection{Mutual information analysis}
\label{subsec-mimethod}
To clarify an indication of the functional differentiation of neurons, we adopted mutual information between neural activity in the internal network and teacher signals.
Mutual information was calculated for spatial and temporal patterns, separately according to the following steps; 
Discretize neuronal states $x_{i}(t)$ by equally dividing the interval between its minimum and maximum values into $n_{d}=8$ states, which are represented by $X_{i}(t) \in \{1,..., n_{d}\}$.
Then, calculate the joint-probability distributions $p(X_{i},l) = p(X_{i}(t) ,l(t-4))$ and
$p(X_{i},m) = p(X_{i}(t) ,m(t-4))$, whose distributions provide the basis of the calculation of mutual information for spatial and temporal patterns defined in the following manner.
\[I^{(\text{sp})}_{i}= -\sum_{X_{i}=1}^{n_d}{ p(X_{i}) \log{(p(X_{i} ))} } +
\sum_{X_{i},l}{p(X_i,l)\log(p(X_i|l)) },\]
\[I^{(\text{temp})}_{i}= -\sum_{X_{i}=1}^{n_d}{ p(X_{i}) \log{(p(X_{i} ))} } +
\sum_{X_{i},m}{p(X_i,m)\log(p(X_i|m)) }.\]

\section{Computation Results}
\label{sec-results}

\subsection{Learning of task by evolution}

\begin{figure}
  \includegraphics[width=8cm]{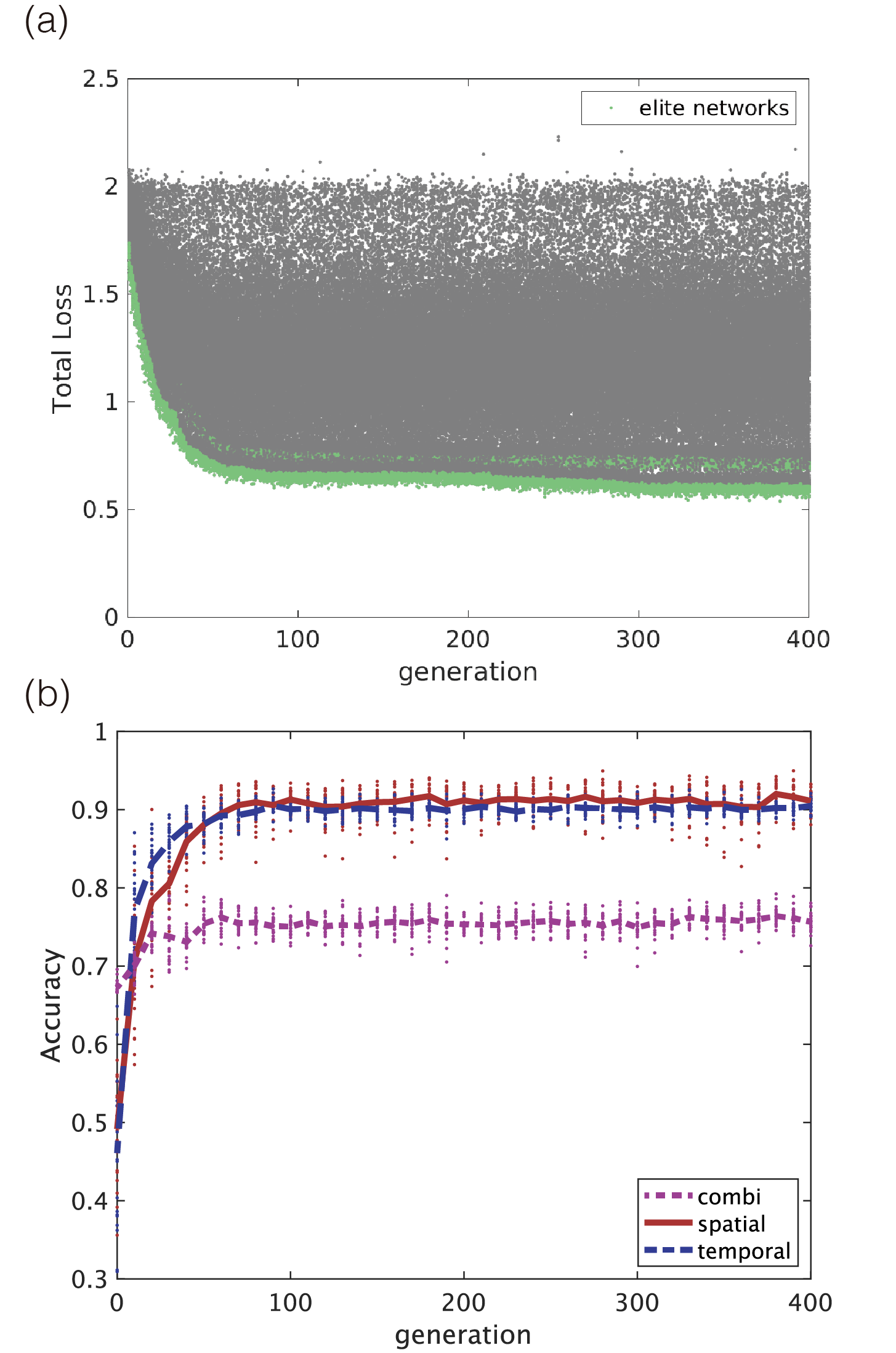}
  \caption{Evolution for a separation task.
  (a) Evolution of loss function. Green dots indicate the loss function of elite individuals that can survive to the next generation, and gray dots indicate other individuals. (b) Evolution of accuracy. Red and blue colors indicate accuracies of spatial and temporal patterns classifications, respectively. Here, only survived networks are shown. The purple color indicates accuracy for the case of the combination task. Because the performance for the combination task is not included in the loss function,
  the performance for the combination task converges to a low level of accuracy. $N=64$.
    \label{fig-evo}
}
\end{figure}

\begin{figure*}
  \includegraphics[width=16.5cm]{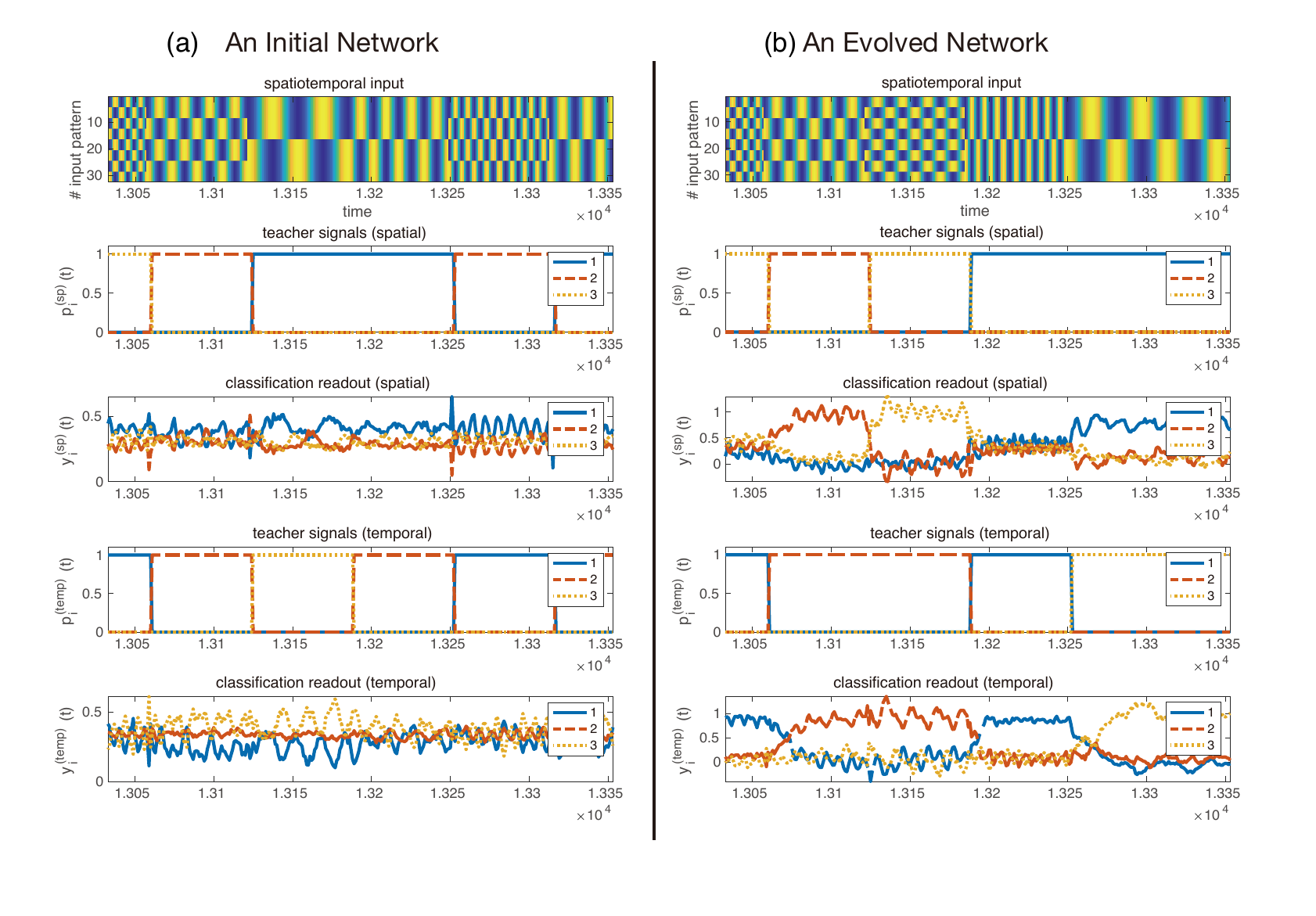} 
    \caption{Example of time series of the network output before and after evolution.
    (a) shows the initial states of a network where learning of only the weights to readout units was performed by ridge regression.
    (b) shows an evolved network. The first row denotes spatiotemporal inputs, the second for spatial teacher signals, the third for spatial readout units, the fourth for
    temporal teacher signals, and the last for temporal readout units. The network learned successfully to output correct signals for both spatial and temporal patterns,
    except for the first few steps after an exchange of the inputs.}
  \label{fig-result-example}

\end{figure*}

  Computation results of evolutionary dynamics showed high accuracy for both spatial and temporal patterns,
which were around 90$\%$ 
in the case that the number of neuron $N$ was $64$ (Fig. \ref{fig-evo}(b)). 
The process of evolution is shown in Fig. \ref{fig-evo}(a). Although the performance of the early networks was low, it rapidly improved during the first hundred generations, thereby converging to a certain equilibrium state.

In this experiment, the objective function depends on the discriminations independently performed for spatial and temporal patterns but independent of the performance of the combination task. Nevertheless, the network performance can be estimated through learning only the readout units for each generation, because
a common feature of RCs is that multiple-readout units can learn in parallel.
In Fig. 2(b), the performance of the network in the case that the combination task was learned by ridge regression for each network, and each generation was also
indicated by the purple dotted line and dots. Because the network evolution was not optimized for the combination task, the network performance had an accuracy of around 70\%.

We show the output time series for spatial and temporal patterns at initial and evolved networks in Fig. \ref{fig-result-example}.
The evolved network successfully outputs correct signals for the classification of spatial and temporal patterns,
except for the first few steps after exchanging the inputs.
Evolutions of spectral radii of the recurrent weight matrix and the distribution of decay constant $\alpha_{i}$ are discussed in Appendix \ref{appendix-additional}.
 
\subsection{Differentiation of neurons and the change of network structure}
\begin{figure}
  \includegraphics[width=8cm]{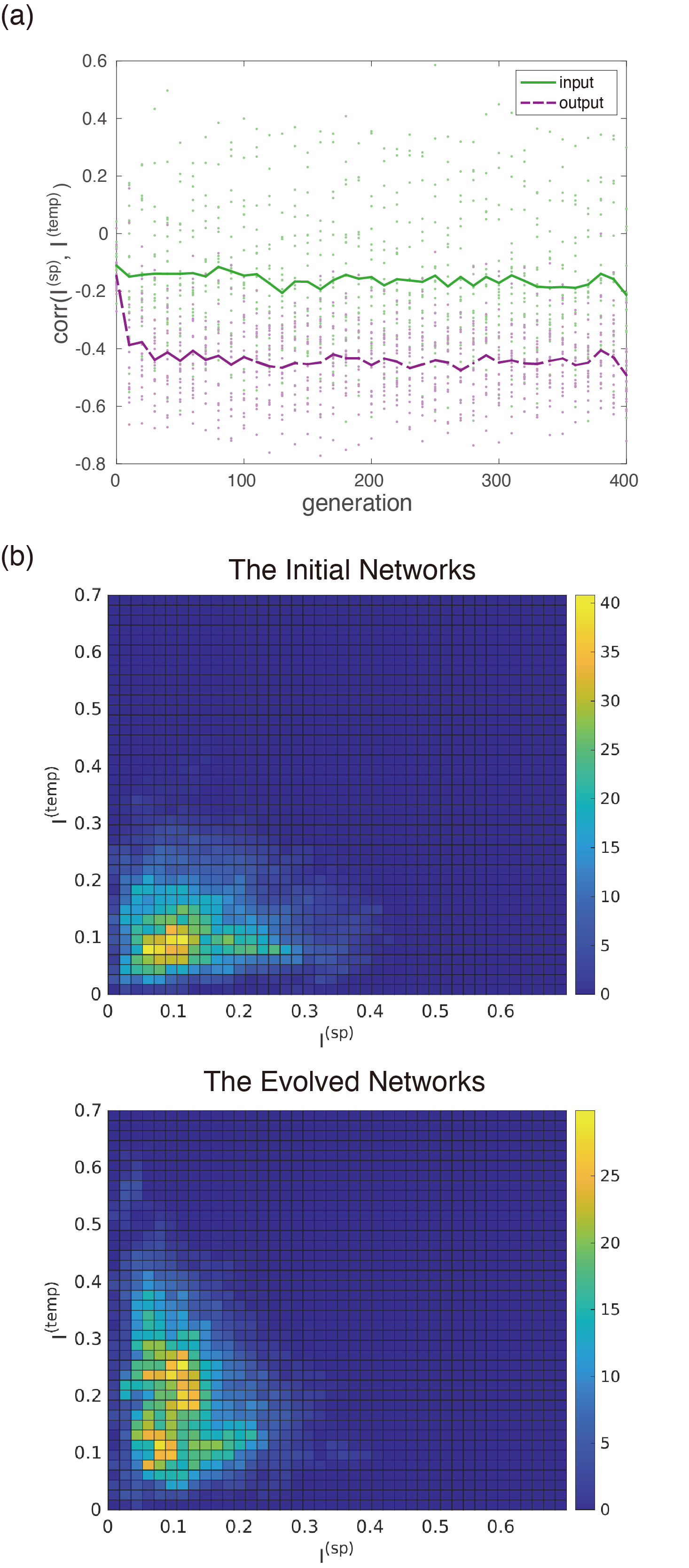}
    \caption{Mutual information analysis. (a) Evolution of correlations between $I^{(\text{sp})}$ and $I^{(\text{temp})}$. Computation results were obtained with 20 different seeds of random numbers. 
    Each dot indicates a top-ranked network at each of the 20 trials, and the lines indicate the averages for all units in the input layer (green color) and in the  output layer (purple color).
    (b) Joint-probability density of $I^{(\text{sp})}$ and $I^{(\text{temp})}$ in the output layer.
     The abscissa denotes $I^{(\text{sp})}$ and the ordinate indicates $I^{(\text{temp})}$ for the initial networks (upper) and for the evolved networks (lower).
    }
  \label{fig-miresult}
\end{figure}

To investigate functional differentiation of each neuron through evolution, we examined mutual information between the state of neuron $x_{i}(t)$ and spatial patterns $l(t)$, and between $x_{i}(t)$ and temporal patterns $m(t)$. 
If two quantities are negatively correlated over neuron ensembles, this implies the existence of many differentiated neurons that primarily represent only one type of information.
The correlation coefficient between the two mutual information values in the input layer was about $-0.1$, indicating little change with evolution, and variance between trials was large (Fig. \ref{fig-miresult}(a)).
In contrast, correlation in the output layer significantly decreased in early evolution stages, when classification performance was improving, finally reaching a trial average of about $-0.4$ (Fig. \ref{fig-miresult}(a)).
Focusing on the output layer that showed a negative correlation, we made histograms to approximate the joint probability distribution of the two mutual information values $I^{\text{(sp)}}$ and $I^{\text{(temp)}}$ in the output layer neurons (Fig. \ref{fig-miresult}(b)).
Compared with the initial networks, the distribution extends to the upper left and lower right after evolution. 
In particular, neurons with higher $I^{\text{(temp)}}$ increased in the evolved networks. 
This result implies the appearance of the functional differentiation of the neural specificity of sensory information.
To test robustness of the emergence of functional differentiation, we conducted two additional experiments: one using spatiotemporal patterns in which every component has a different oscillatory phase and another using a different set of frequencies. Appendix \ref{appendix-additional} presents the results.

\begin{figure}
  \includegraphics[width=8cm]{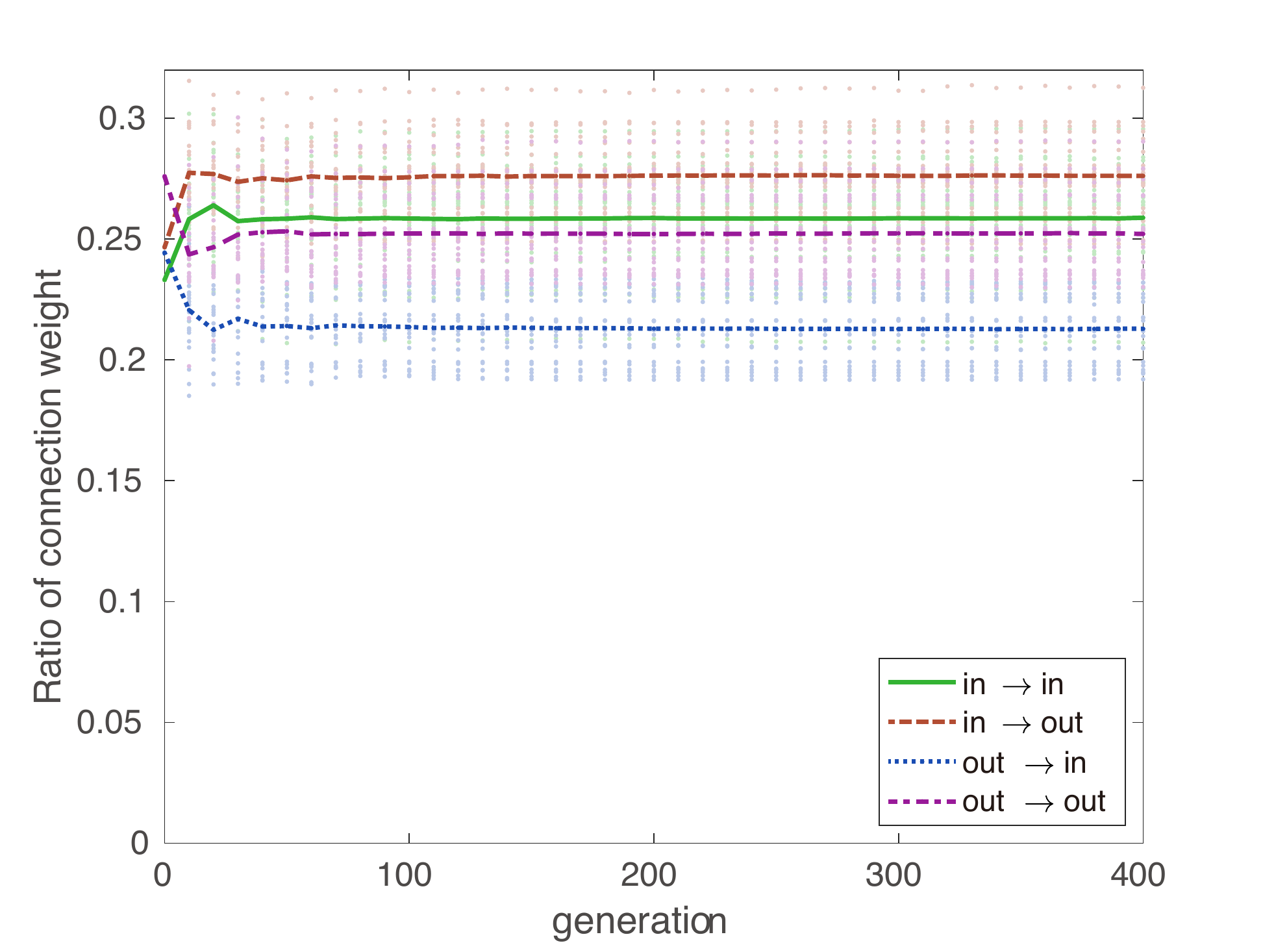}
    \caption{  The change of network structure in the process of evolution.
    The ratios of connection weights inside and between layers are shown. Colored lines indicate the averaged values over 20 trials with different seeds. 
    Absolute values of weights were considered, so that the ratio of excitatory and inhibitory connections is not considered. 
    }
  \label{fig-ratio-connection}
\end{figure}

We also investigated how the network structure changed from a random network after evolutionary optimization was established.
We found that the feedforward connections from the input layer to the output layer in the internal network were
strengthened much more than the other connections (Fig. \ref{fig-ratio-connection}). Furthermore, the feedback connections  from the output to the input layer were rather weakened, and were
also weakened for internal feedback connections in both the input and output layers.
We estimated the effective connectivity of the network type for the feedforward or feedback type of networks
by such connection strength.
The ratio of effective connectivity of the feedback to the feedforward networks varied depending on the system size $N$: about 2:3 at $N=32$, and 3:4-- 6:7 at $N=64$.
This tendency of the reduction of feedback connections relative to the feedforward connections
does not contradict the structure of cortical local connections that was recently found by Seeman et al.\cite{seeman2018}, where
the ratio of the number of connections was reported as 1:10 for mouse primary visual cortex and 1:5-- 1:7 for
human frontal and temporal cortices.
Furthermore, the evolved structure shown in the present ERC is similar to the evolutionary change of the network structure
of hippocampus. In the biological evolution from reptiles to mammals, the hippocampus evolved from rather randomly uniform  networks consisting of a small-cell and a large-cell layer to differentiated feedforward networks including feedback connections
\cite{treves2004}. It is also  known in mammals that some amount of feedback connections exist in CA3, whereas only a few feedback connections exist in CA1. Thus, the ratio of feedback connections to feedforward connections in the mammalian hippocampus
looks consistent with our findings in ERCs.
However, it is questionable if the constraint adopted here was used in biological evolution. In biological evolution,
sexual and social constraints as well as natural selections would have been more significant than the information constraint adopted here.

The structural changes described may induce functional changes. 
It is known that uniformly random networks in reptiles give rise to a simple memory, which can be realized by a single association. In contrast, 
RNNs with inhibitory neurons in mammals 
can produce an episodic memory, which can be realized by a successive association of memories (see, for example,
 \cite{tsuda1987,tsukada2013}). The hippocampal CA3 plays a role in the formation of such sequences of memories, while 
 the hippocampal CA1 receives such sequences, and 
 encodes the sequential information in a form of fractal geometry, called 
 Cantor coding (see, for example, \cite{tsuda2001,fukushima2007,tsuda-kuroda2001,kuroda2009,yamaguti2011}). 
 Considering these findings,  the evolution of the 
 network structure to the feedforward network including feedback connections was a decisive milestone in the evolution of memory. 
 Therefore, the present ERC possesses sufficient structures of neural networks to yield high performance for not only the discrimination of patterns but also memory formation.

\subsection{Combination task}
\label{sec-combinationtask}

\begin{figure}
  \includegraphics[width=8cm]{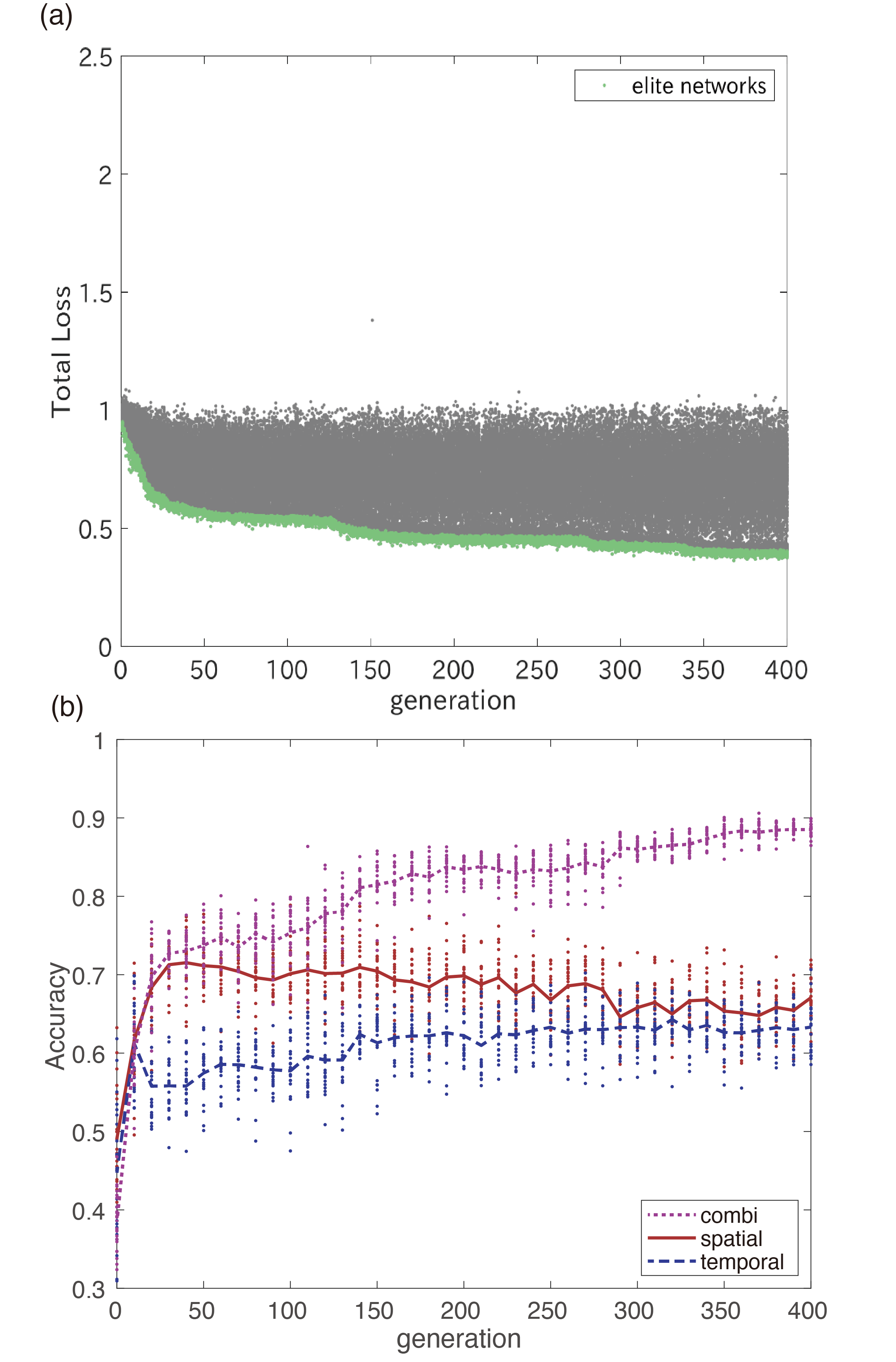}
  \caption{Example of accuracy changes when the network evolved to optimize for the combination task.
  (a) Evolution of the values of the loss function. Green dots indicate elite networks that survived in the next generation, and gray dots indicates other networks. 
  (b) Accuracy changes. Purple denotes the accuracy for the combination task, red denotes the accuracy for spatial patterns, and blue denotes the accuracy for temporal patterns. Only survived elite networks are shown. The combination task accuracy increases up to 90\%, but both spatial and temporal patterns are not so recognized. $N=64$.
    }
  \label{fig-evo-combi}
\end{figure}
\begin{figure*}
  \includegraphics[width=17cm]{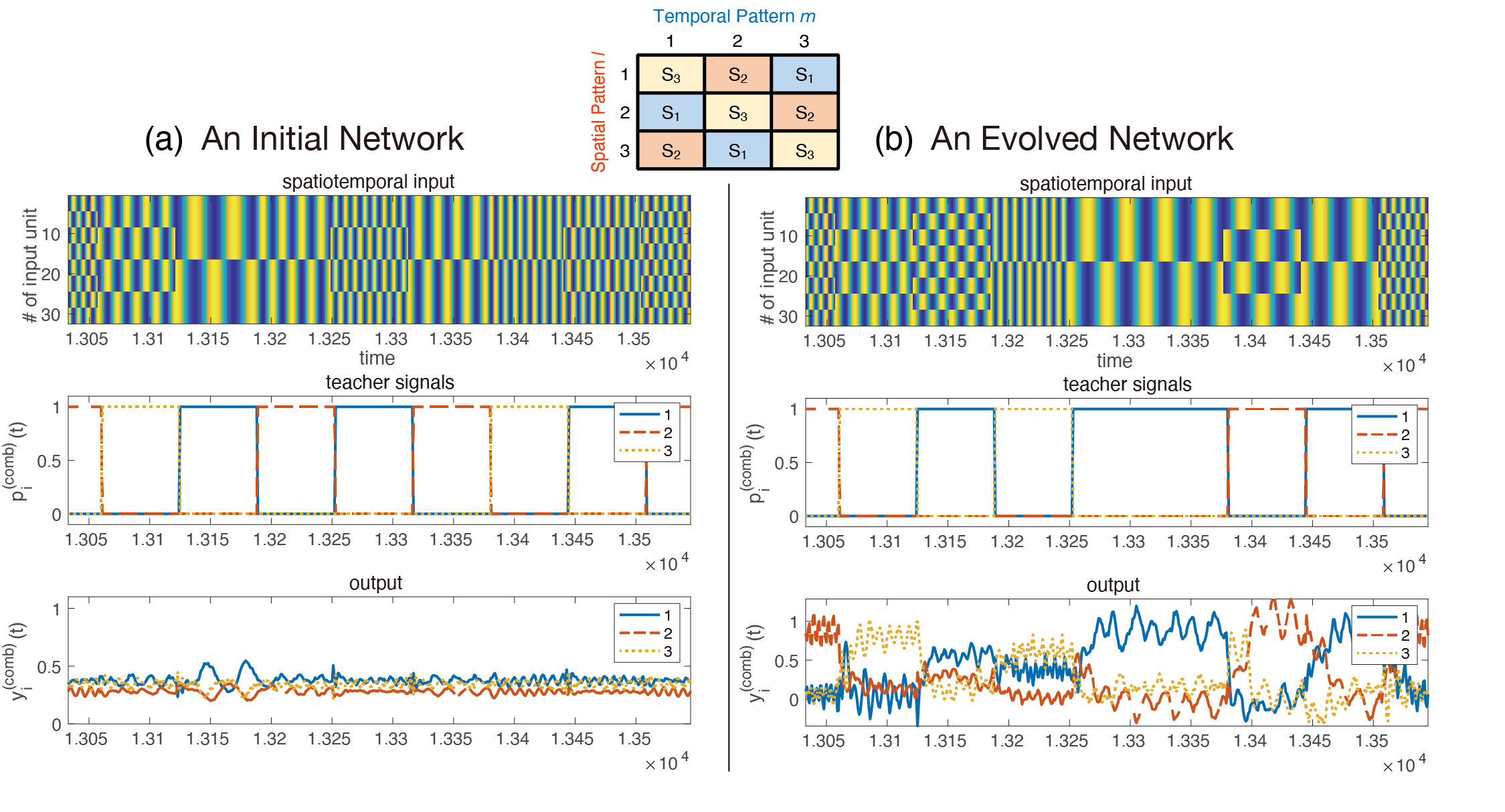}
  \caption{Example of a time series of network outputs before ((a)) and after ((b)) evolution in the case of the combination task. 
  The top figure indicates the allocation of combinations of spatial and temporal patterns to readout units.
  The time series of initial networks is shown in (a), and in (b) evolved network is shown. An initial network means a random network where only readout weights were trained by ridge regression. The first row indicates spatiotemporal inputs, the second indicates teacher signals for combination patterns, and the third indicates readout unit activity. 
  The network successfully learned teacher signals after evolution.
  }
    \label{fig-ex-combi}
\end{figure*}

\begin{figure}
  \includegraphics[width=8cm]{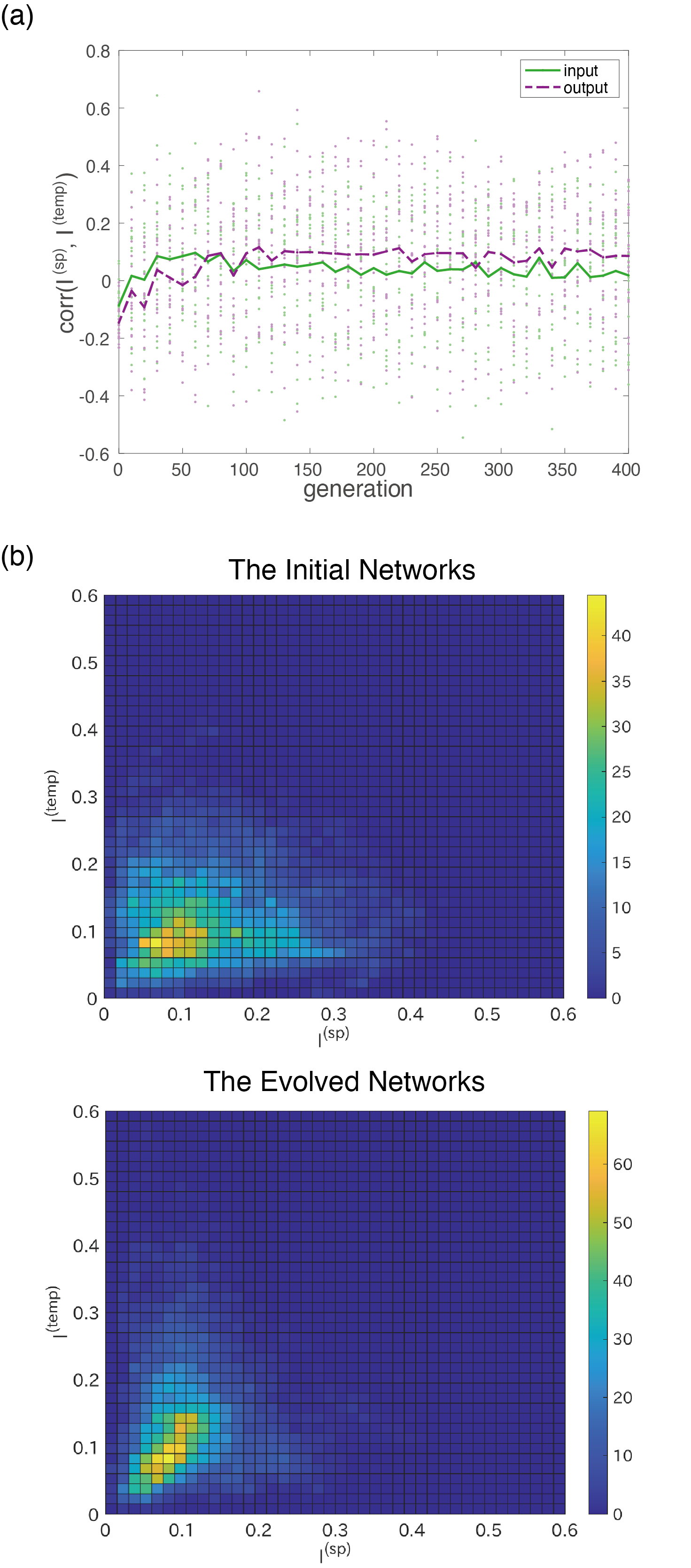}
    \caption{Mutual information analysis of the network optimized for the combination task.
    (a) Evolution of correlations between $I^{(\text{sp})}$ and $I^{(\text{temp})}$. 
    Computation results were obtained with 20 different seeds of random numbers. 
    Each dot indicates the top network at each of 20 trials and lines for the averages 
    for units in the input layer (green color) and in the output layer (purple color). 
    (b) Joint-probability density of $I^{(\text{sp})}$ and $I^{(\text{temp})}$ in the output layer. The abscissa denotes the $I^{(\text{sp})}$ and the ordinate denotes the $I^{()\text{temp})}$ for the initial networks (upper) and for the evolved networks (below). 
      }
  \label{fig-miresult-combi}
\end{figure}

For comparison with the presented task, called here the separation task, 
we investigated the development of networks in the case of the combination task, where the same network size and parameters were used. 
The combination task identifies specific combinations of the spatial and temporal patterns.
Learning succeeded with a high accuracy (Fig. \ref{fig-evo-combi}, \ref{fig-ex-combi}), but 
a similar differentiation was not observed in the internal network structure,
that is, $I^{(\text{sp})}$ and $I^{(\text{temp})}$ did not show negative correlations(Fig. \ref{fig-miresult-combi}(a)).
 The joint probability distribution of the evolved networks (Fig. \ref{fig-miresult-combi}(b)) shows a concentration along a diagonal line, which was not seen in the separation task.
This implies that a different type of optimization occurred. Specifically, because many neurons possessing both spatial and temporal information were observed in the output layer, such neurons possibly do not separate sensory information, but rather unify sensations, thereby processing a combination of sensory inputs.
The emergence of such neurons may be related to the concept of ``mixed selectivity,'' which has recently been intensively studied both experimentally and theoretically in the  neocortex \cite{rigotti2013}. 

\subsection{Dependence on constraints}

\begin{figure} 
  \includegraphics[width=8cm]{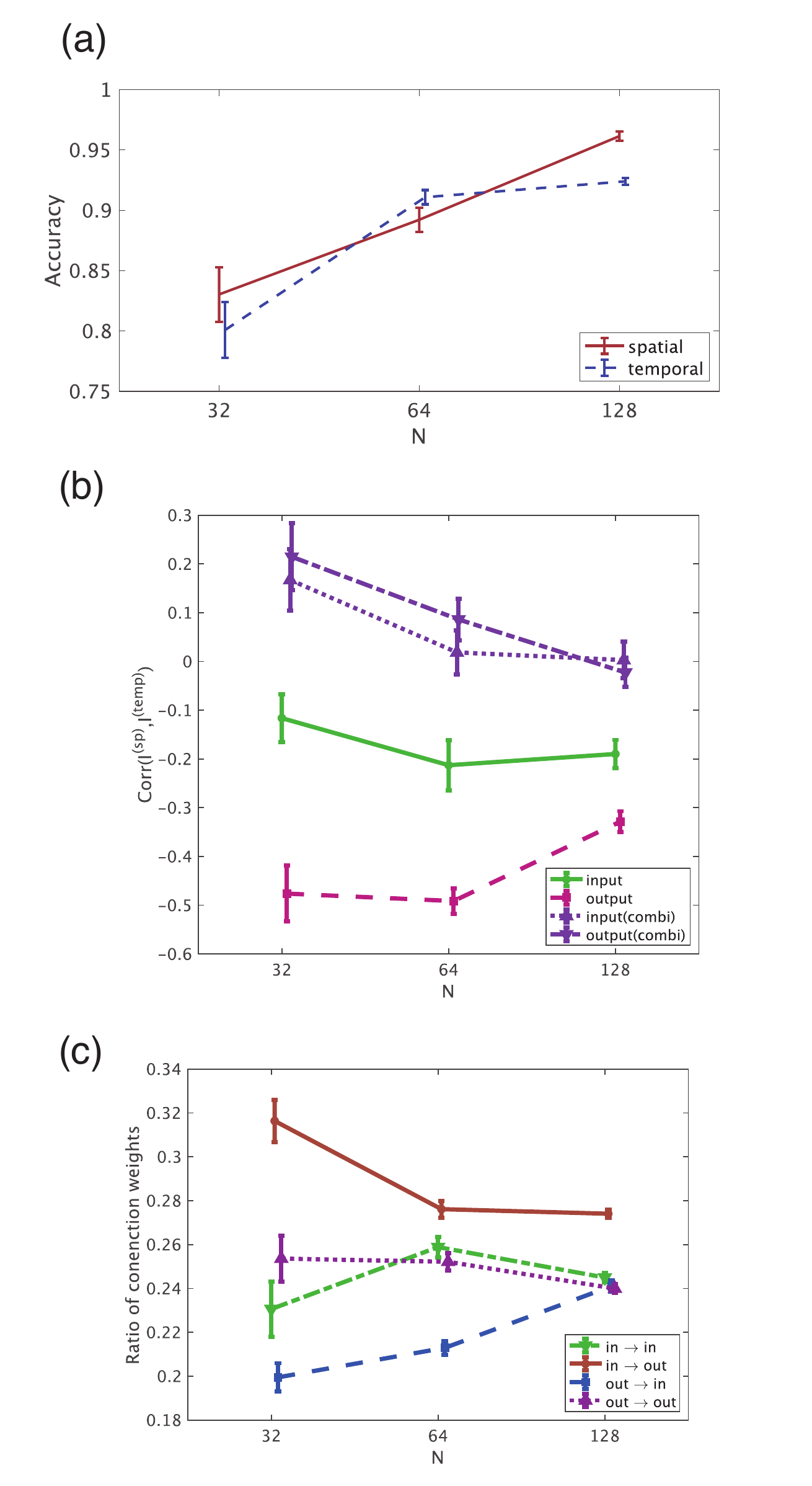}
  \caption{Dependence of differentiation on network size.
  (a) Accuracy, (b) correlation between $I^{(\text{sp})}$ and $I^{(\text{temp})}$, and 
  (c) ratio of connections. In all figures, error bars mean standard error of the mean for 20 trials. 
  }
  \label{fig-dependence-n}
\end{figure}

\begin{figure}
  \includegraphics[width=8cm]{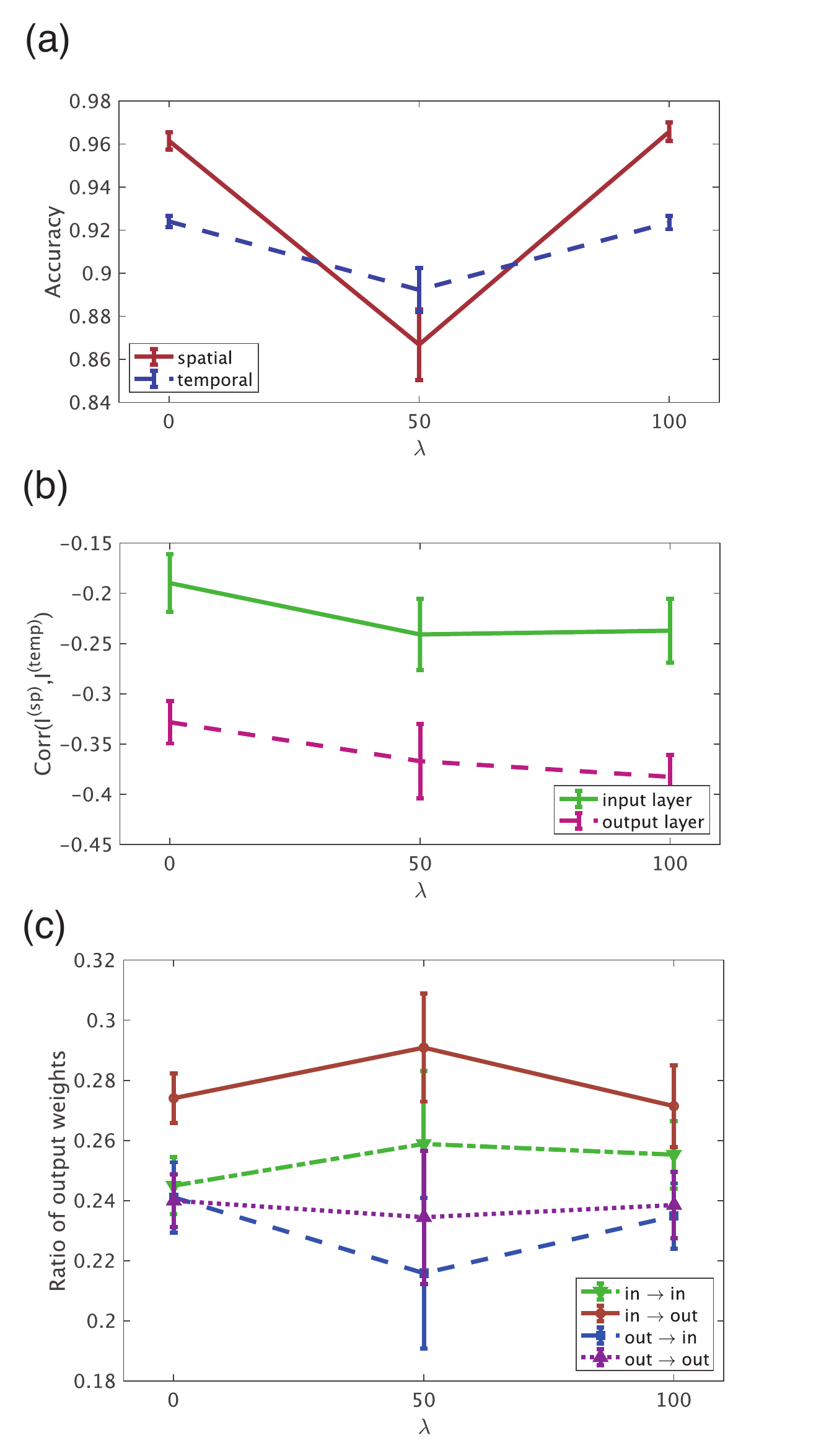}
  \caption{
   Dependence of differentiation on weight constraints $\lambda$ in the case of $N = 128$.
  (a) Accuracy, (b) correlation between $I^{(\text{sp})}$ and $I^{(\text{temp})}$, and
  (c) ratio of connections.  In all figures, error bars mean standard error of the mean for 20 trials.
  }
  \label{fig-dependence-l2}
\end{figure}

We investigated two kinds of conditions on which differentiation depends: the size of the network and the strength of the weights.
As for the network size, the computation results showed that a decreasing network size corresponds to an increasing number of 
neurons that specifically respond to either input signal of spatial or temporal patterns (Fig. \ref{fig-dependence-n}). 
Classification accuracy increased with network size (Fig. \ref{fig-dependence-n}(a)). 
However, the degree of functional differentiation in the output layer, as measured by the strength of negative correlation of mutual information, was higher at $N=32$ and $N=64$ than at $N=128$ (Fig. \ref{fig-dependence-n}(b)). 
There was no negative correlation in any case in the combination task. 
As for network structure, we found that ratios of feedforward connections increased and those of feedback connections decreased as the network size decreased (Fig. \ref{fig-dependence-n}(c)).

To investigate the effect of penalty for weight size, the following penalty term,
$\lambda \sum_{ij} {w_{ij}^{2}/N^2}$, on the connections of the internal network was added to the loss function and evolved the network.
Here, we show the computation results in the case of $N=128$. Because the network with $N=64$ sufficiently differentiates 
even without such an additional penalty term, the effect of the additional term was very weak. In contrast, in $N=128$,
because the degree of differentiations was low in the case without the penalty term (Fig. \ref{fig-dependence-n}(b), we investigated how differentiations changed by such an additional term (Fig. \ref{fig-dependence-l2}). 
As the strength of the penalty $\lambda$ increased, accuracy fell at $\lambda=50$ and rose again at $\lambda=100$ (Fig. \ref{fig-dependence-l2}(a)). A detailed investigation of the reasons for this decrease and increase is beyond the scope of this paper.
Negative correlations between two mutual information values became stronger as $\lambda$ increased, implying an increase in differentiated neurons (Fig. \ref{fig-dependence-l2}(b)).
The feedforward-to-feedback ratio did not strongly depend on the penalty strength (Fig. \ref{fig-dependence-l2}(c)). 
  It looks the penalty strength did not affect the differentiation on feedforward/feedback structure of the network.
In summary, there was an evident range of penalty strengths that enhances differentiation without adversely affecting classification accuracy.

\section{Discussion}
\label{sec-discussion}
In this paper, we proposed a new computation network type that simultaneously decodes multiple kinds of inputs, using the
evolution of an internal network of RCs. The network performance studied with 
mutual information analysis showed neuronal differentiations in the sense of the emergence of  the 
specificity of neurons, responding to a single specific input pattern. In the simulation, some other neurons still evolved, 
responding to multiple inputs. This type of neuron seems capable of evolving to neurons that may specifically respond to other new kinds of inputs. The computation results suggest the information mechanism underlying 
functional differentiations in the brain. 

We also investigated the dynamics of the evolved networks. After the optimization is established, all evolved networks 
produced a single fixed-point attractor in the case without inputs. In the presence of input sequence, a limit-cycle type of quasi-attractors emerged and transitions between such quasi-attractors occurred. These dynamic behaviors appear 
different from chaotic itinerancy \cite{kanekotsuda2003}, which represents chaotic transitions between quasi-attractors.

\begin{acknowledgments}
  The authors would like to thank Otto E. R\"ossler for his continual encouragement and stimulation via both personal communication and 
  his outstanding work on chaos. We would like to thank Y. Kawai, J. Park, M. Asada, T. Hirakawa, T. Yamashita, H. Fujiyoshi, and H. Watanabe for their helpful discussions.
  This study was partially supported by the JST Strategic Basic Research Programs (Symbiotic Interaction: Creation and Development of Core Technologies Interfacing Human and Information Environments, CREST Grant Number JPMJCR17A4).
  This work was also partially supported by a Grant-in-Aid for Scientific Research on Innovative Areas (Non-Linear Neuro-Oscillology: Towards Integrative Understanding of Human Nature, KAKENHI Grant Number 15H05878) from the Ministry of Education, Culture, Sports, Science and Technology, Japan.
  \end{acknowledgments}

  \appendix
\section*{Data Availability}
The data that supports the findings of this study are available within the article.

\section{Parameter values}
Here, we describe parameter values used in numerical simulations of this article. The number of the 
reservoir units: $N=64$, except for Fig. \ref{fig-dependence-l2} where $N=128$. 
Both the number of neurons in the input layer $N_{\text{in}}$ and the number of input units are $N/2$.
The size of
the population in the gene pool: $N_{pop}=220$. The number of survived elite networks in each 
 generation: $N_{\text{suv}}=22$, the number of new networks generated by mutation: $N_{\text{mut}}=128$, and for the networks generated by crossover: $N_{\text{cross}}=72$. The probability of an
occurrence of  the recombination of connections during mutation is 0.04. The probability of changing
connection weights during mutation is 0.4. The standard deviation of perturbation for weights is
 $\sigma_{w} = 0.05$. The probability of an occurrence of perturbation for the decay constant
  $\alpha_i$ during mutation is 0.1, whose standard deviation $\sigma_{\alpha}$ is 0.01.

For each network in each generation, $T_{\text{tr}}=12000$ steps after the first $1000$ steps of transient
were used for learning, in which we recorded the internal states
and the teacher readout signals, and then determined the weights 
$w^{(\text{sp})}$ and $w^{(\text{temp})}$ such as minimizing the loss function in Eq.~(\ref{eq-loss}) 
using a ridge regression method. Then, further $T_{\text{te}}=10000$ 
steps were used for the test, in which we evaluated the difference between the teacher signal 
and the readout state $y$, and determined the values of the objective function for evolution, following Eq.~(\ref{eq-objective}).

At the beginning of the evolution, a sparse recurrent weight matrix $W=(w_{ij})$ was randomly created, and their weights were rescaled such that the spectral radius of the matrix became one. The values of the input weights $w^{(\text{in})}_{ik}$ took 0.1 if $i=k$ and zero otherwise. The noise term $\xi$ was given by a normal distribution of mean $0$ and a standard deviation of $0.001$.

\section{Additional analysis and experiments}
\label{appendix-additional}
In this section, we describe additional analyses and experiments.

\subsection{Evolution of the spectral radius}

\begin{figure} 
  \includegraphics[width=6cm]{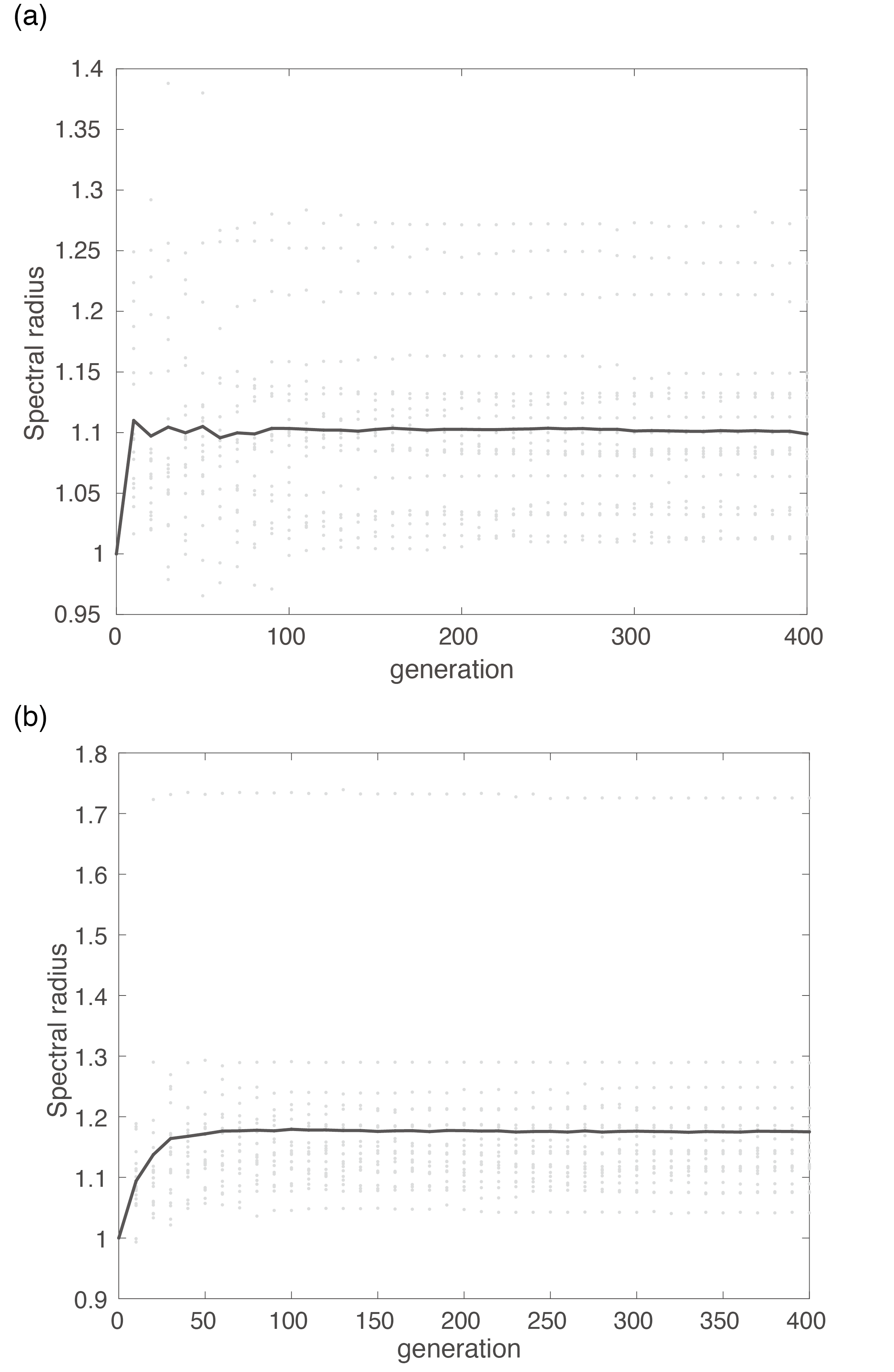}
  \caption{Evolution of spectral radius of the recurrent weight matrix. (a) The separation task. (b) The combination task. Each dot indicates the top network at each of 20 trials and the solid lines indicate the averages. }
  \label{fig-spectral-radius}
\end{figure}

To investigate the evolution of overall connection strengths that may affect network dynamics, we observed evolution of the spectral radius of the recurrent weight matrix (Fig. \ref{fig-spectral-radius}).
In both the separation and the combination task, averages of spectral radii exceeded 1.0 after evolution. 
The spectral radius was higher in the combination task than in the separation task, indicating different optimal connection strengths between the two tasks.
Nevertheless, in all cases, fixed-point attractors were observed when we ran the evolved networks with no input (figures not shown).


\subsection{Evolution of the decay constant}

\begin{figure} 
  \includegraphics[width=8cm]{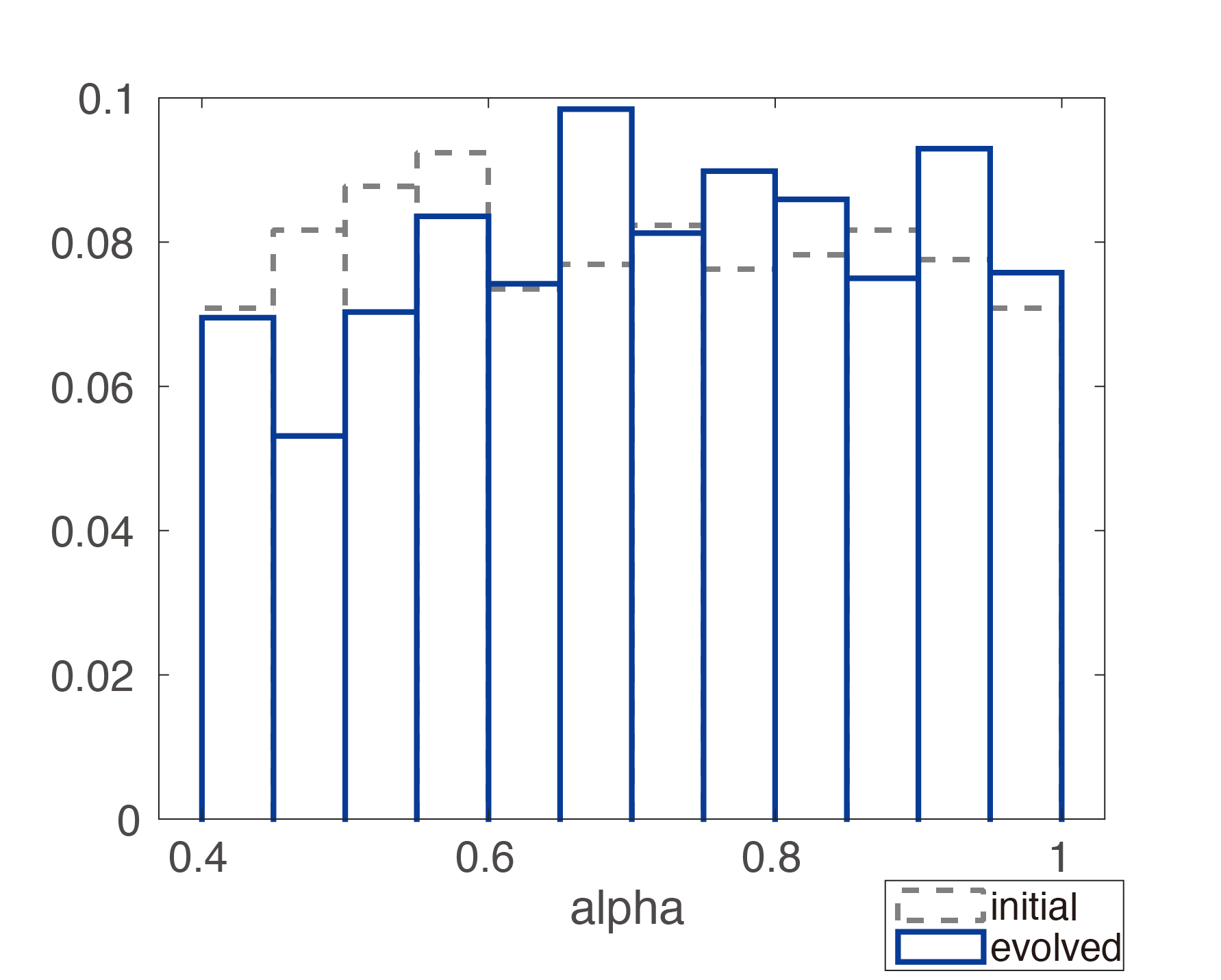}
  \caption{The distribution of the decay constant $\alpha_{i}$.}
  \label{fig-alpha}
\end{figure}

  In our genetic algorithm, the decay constant of each neuron was also subject to evolution. To determine whether decay constants adapted to the time scale of the input signal through evolution,  we studied changes in the distribution of the decay constant $\alpha_{i}$ (Fig. \ref{fig-alpha}). This distribution did not change much from the initial network to the evolved network. The question of what information is encoded by a neuron with specific time constants is left for future research.


\subsection{Temporal pattern with phase shift}
\begin{figure} 
  \includegraphics[width=8cm]{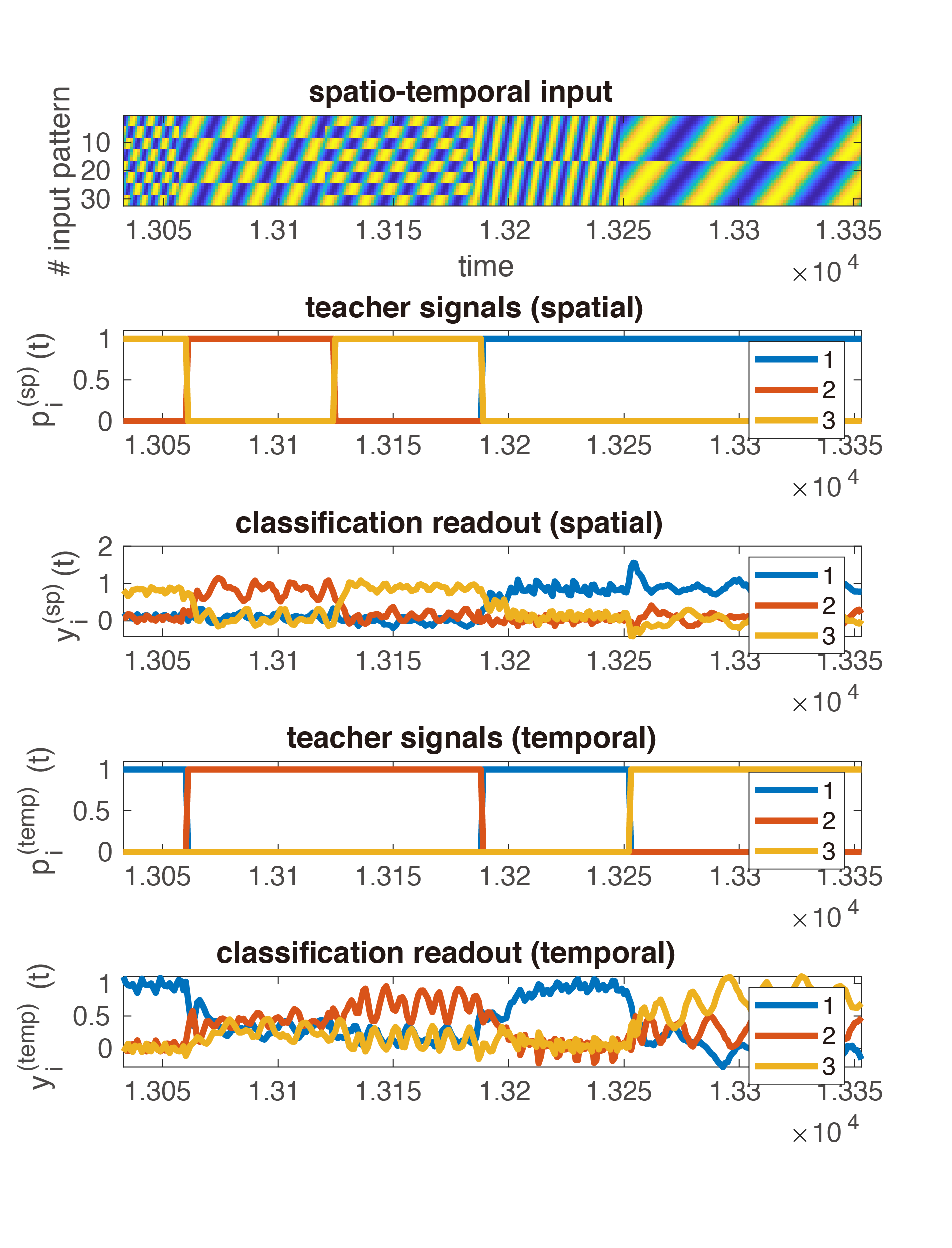}
  \caption{Example of time series of input and output of the network evolved with phase-shifted spatio-temporal inputs.
  The first row denotes spatiotemporal inputs, the second for spatial teacher signals, the third for spatial readout units, the fourth for temporal teacher signals, and the last for temporal readout units. }
  \label{fig-phaseshift-ex}
\end{figure}

\begin{figure} 
  \includegraphics[width=8cm]{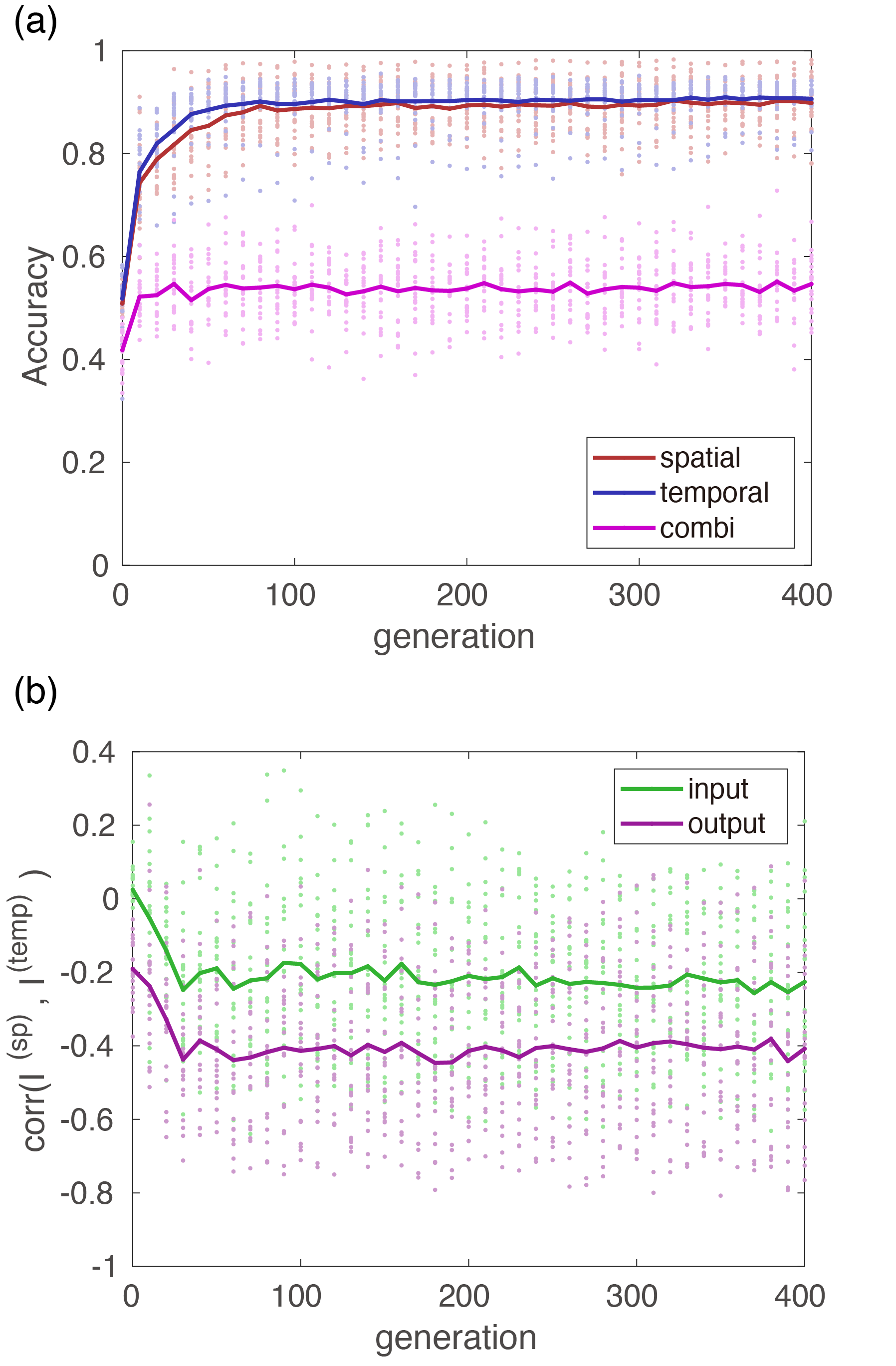}
  \caption{Evolution of the network with phase-shifted spatio-temporal inputs.  All parameters are same as in Fig. \ref{fig-evo}-\ref{fig-miresult}, except for input signals. Computation results were obtained with 20 different seeds of random numbers. (a) Evolution of accuracy.  Solid lines indicate the averages of 20 trials. Red and blue colors indicate accuracies of spatial and temporal patterns classifications, respectively.  The purple color indicates accuracy for the case of the combination task.
  (b) Evolution of correlations between $I^{(\text{sp})}$ and $I^{(\text{temp})}$. 
  Each lines indicate the averages for all units in the input layer (green color) and in the  output layer (purple color).}
  \label{fig-phaseshift-sum}
\end{figure}

    To test the robustness of emergence of the functional differentiation, we conducted an additional evolution experiment in which input signal elements had mutually different phases. Specifically, the temporal patterns were modified to 
          \[ b^{(\bar{m})}_{k}(t)=\cos\left(2 \pi (f^{(\bar{m})} t + k/N_{\text{in}})\right),\]
    where $k=1,2,..., N_{\text{in}}$ and $\bar{m}=1, ..., M$.
    Then the $k$-th component of the input signal was 
          \[I_k(t)  = a^{(l(t))}_{k} b^{(m(t))}_{k}(t). \]
    Figure \ref{fig-phaseshift-ex} shows examples of spatiotemporal input patterns, teacher signals, and readouts of the evolved network. Performance of the evolved network was at the same level as that in the main results (Fig. \ref{fig-phaseshift-sum}(a)). The degree of differentiation, as measured by the strength of negative correlation of mutual information, was slightly higher than that in the main results (Fig. \ref{fig-phaseshift-sum}(b)).
    In summary, the network showed the same evolutionary adaptability and differentiation as with the standard conditions, even when there was a phase difference in the input.

\subsection{Temporal pattern with different periods}

\begin{figure} 
  \includegraphics[width=8cm]{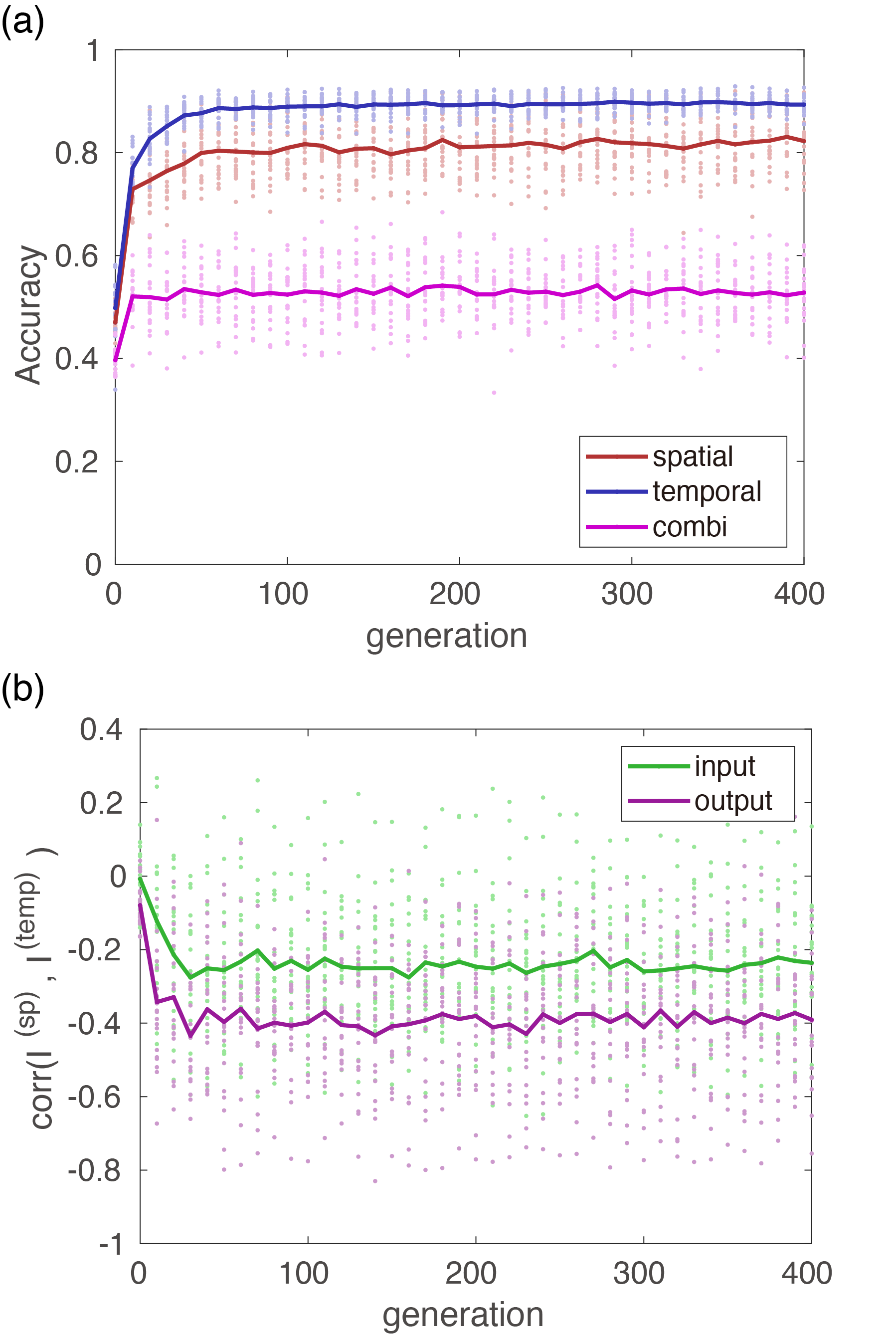}
  \caption{Evolution of the network with the different set of frequencies $\{f^{(m)}\}=\{1/7,1/19,1/31\}$.   (a) Evolution of accuracy.  
  (b) Evolution of correlations between $I^{(\text{sp})}$ and $I^{(\text{temp})}$. 
  Meanings of styles and colors of lines and dots are the same as the previous figure.}
  \label{fig-diffreqs}
\end{figure}

To further test robustness of the functional differentiation against change in frequency, we performed evolution experiments with a different frequency set $\{f^{(\bar{m})}\}=\{1/7,1/19,1/31\}$, whose inverses (oscillation periods) are prime numbers.
   Other parameters values were the same as in the other simulations. 
   Under these conditions, the classification accuracy of spatial patterns decreased to about $85\%$ (Fig. \ref{fig-diffreqs}(a)), but the degree of differentiation did not change (Fig. \ref{fig-diffreqs}(b)). 
   This degradation in spatial pattern classification might be related to sinusoidal waveform distortions due to coarse sampling using odd periods.

\bibliography{refers_chaos}
\end{document}